\documentclass[10pt]{article}

\usepackage{amsmath}
\usepackage{amssymb}
\usepackage{graphics}

\title{\textbf{Quantum gravity computers: On the theory of computation with indefinite causal structure}}

\author{Lucien Hardy\\
\textit{Perimeter Institute,}\\
\textit{31 Caroline Street North,}\\
\textit{Waterloo, Ontario N2L 2Y5, Canada}}

\begin{document}

\maketitle

\begin{abstract}

A quantum gravity computer is one for which the particular effects of quantum gravity are relevant. In general
relativity, causal structure is non-fixed. In quantum theory non-fixed quantities are subject to quantum uncertainty.
It is therefore likely that, in a theory of quantum gravity, we will have indefinite causal structure. This means that
there will be no matter of fact as to whether a particular interval is timelike or not. We study the implications of
this for the theory of computation. Classical and quantum computations consist in evolving the state of the computer
through a sequence of time steps.  This will, most likely, not be possible for a quantum gravity computer because the
notion of a time step makes no sense if we have indefinite causal structure. We show that it is possible to set up a
model for computation even in the absence of definite causal structure by using a certain framework (the causaloid
formalism) that was developed for the purpose of correlating data taken in this type of situation.  Corresponding to a
physical theory is a causaloid, ${\bf\Lambda}$ (this is a mathematical object containing information about the causal
connections between different spacetime regions).  A computer is given by the pair $\{ {\bf\Lambda}, S\}$ where $S$ is
a set of gates. Working within the causaloid formalism, we explore the question of whether universal quantum gravity
computers are possible. We also examine whether a quantum gravity computer might be more powerful than a quantum (or
classical) computer. In particular, we ask whether indefinite causal structure can be used as a computational resource.
\end{abstract}

\section{Introduction}

A computation, as usually understood, consists of operating on the state of some system (or collection of systems) in a
sequence of steps. Turing's universal computer consists of a sequence of operations on a tape.  A classical computation
is often implemented by having a sequence of operations on a collection of bits and a quantum computation by a sequence
of operations on a collection of qubits.  Such computations can be built up of gates where each gate acts on a small
number of bits or qubits. These gates are defined in terms of how they cause an input state to be evolved.  A physical
computer may have some spatial extension and so gates may be acting at many different places at once. Nevertheless, we
can always foliate spacetime such that we can regard the computer as acting on a state at some time $t$ and updating it
to a new state at time $t+1$, and so on, till the computation is finished. Parallel computation fits into this paradigm
since the different parts of the parallel computation are updated at the same time. The notion that computation
proceeds by a sequence of time steps appears to be a fairly pervasive and deep rooted aspect of our understanding of
what a computation is. In anticipation of more general computation, we will call computers that implement computation
in this way {\it step computers} (SC). This includes Turing machines and parallel computers, and it includes classical
computers and quantum computers.

Turing developed the theory of computation as a formalization of mathematical calculation (with pencil, paper, and
eraser for example)\cite{Turing}. Deutsch later emphasized that any computation must be implemented physically
\cite{Deutsch}. Consequently, we must pay attention to physical theories to understand computation. Currently, there
are basically two fundamental physical theories, quantum theory (QT) and Einstein's theory of general relativity (GR)
for gravity. However, we really need a physical theory which is more fundamental - a theory of quantum gravity (QG).  A
correct theory of QG will reduce to QT and GR in appropriate situations (including, at least, those situations where
those physical theories have been experimentally verified). We do not currently have a theory of quantum gravity.
However, we can hope to gain some insight into what kind of theory this will be by looking at QT and GR.  Causal
structure in GR is not fixed in advance.  Whether two events are time-like or not depends on the metric and the metric
depends on the distribution of matter.  In quantum theory a property that is subject to variation is also subject to
quantum uncertainty - we can be in a situation where there is no matter of fact as to the value of that quantity. For
example, a quantum particle can be in a superposition of being in two places at once.  It seems likely that this will
happen with causal structure.  Hence, in a theory of QG we expect that we will have indefinite causal structure.
\begin{quote}
{\bf Indefinite causal structure} is when there is, in general, no matter of fact as to whether the separation
between two events is time-like or not.
\end{quote}
If this is, indeed, the case then we cannot regard the behaviour of a physical system (or collection of systems) as
evolving in time through a sequence of states defined on a sequence of space-like hypersurfaces. This is likely to have
implications for computer science. In particular, it is likely that a quantum gravity computer cannot be understood as
an instance of a SC.  In this paper we will explore the consequences of having indefinite causal structure for the
theory of computation.  In particular, we will look at how the causaloid framework (developed in \cite{Hardy1}) can be
applied to provide a definite model for computation when we have indefinite causal structure. Although there are
compelling reasons for believing that the correct theory of QG will have indefinite causal structure, it is possible
that this will not be the case.  Nevertheless, in this paper we will assume that QG will have this property.  There may
be other features of a theory of QG which would be interesting for the study of computation but, in this paper, we will
restrict ourselves to indefinite causal structure.

\section{General ideas}

\subsection{What counts as a computer?}\label{whatcounts}

The idea of a computer comes from attempting to formalize mathematical calculation.  A limited notion of computation
would entail that it is nothing more than a process by which a sequence of symbols is updated in a deterministic
fashion - such as with a Turing machine.  However, with the advent of quantum computation, this notion is no longer
sufficient. David Deutsch was able to establish a theory of quantum computation which bares much resemblance to the
theory of classical computation. Given that quantum computers can be imagined (and may even be built one day) we need a
richer notion of computation.  However, a quantum computer still proceeds by means of a sequence of time steps.  It is
a SC.  The possibility of considering time steps at a fundamental level will, we expect, be undermined in a theory of
quantum gravity for the reasons given above.

This raises the question of whether or not we want to regard the behaviour of a physical machine for which the
particular effects of QG are important and lead to indefinite causal structure as constituting a computer.  We
could certainly build a machine of this nature (at least in principle). Furthermore, somebody who knows the laws by
which this machine operates could use it to address mathematical issues (at the very least they could solve
efficiently the mathematical problem of generating numbers which would be produced by a simulation of this machine
in accordance with the known laws). Hence, it is reasonable to regard this machine as a computer - a quantum
gravity computer.

At this point it is worth taking a step back to ask, in the light of these considerations, what we mean by a the
notion of a computer in general?  One answer is that
\begin{quote}
(1) {\bf A computer} is a physical device that can give correct answers to well formulated questions.
\end{quote}
For this to constitute a complete definition we would need to say what the terms in this definition mean. However,
whatever a \lq\lq well formulated question" means, it must be presented to the computer in the form of some physical
input (or program). Likewise, whatever an \lq\lq answer" is, it must be given by the computer in the form of some
physical output.   It is not clear what the notion of \lq\lq correctness" means. However, from the point of view of the
physical computer it must mean that the device operates according to sufficiently well known rules.  Hence, a more
physical definition is that
\begin{quote}
(2) {\bf A computer} is a physical device has an output that depends on an input (or program) according to
sufficiently well known rules.
\end{quote}
This still leaves the meaning of the word \lq\lq sufficiently" unclear. It is not necessary that we know all the
physics that governs a computer.  For example, in a classical computer we do not need to have a detailed understanding
of the physics inside a gate, we only need an understanding of how the gate acts on an input to produce an output.
There remain interesting philosophical questions about how we understand the translation from the terms in definition
(1) to those in definition (2) but these go beyond the scope of this paper.

These definitions are useful. In particular they do not require that the computational process proceed by a sequence of
steps.  We will see how we can meaningfully talk about computation in the absence of any spacelike foliation into
timelike steps in the sense of definition (2) of a computer.

It is likely that, in going to QG computers, we will leave behind many of the more intuitive notions of computation we
usually take for granted.  This already happened in the transition from classical to quantum computation - but the the
likely failure of the step computation model for a QG computer may cause the transition from quantum to quantum gravity
computation to be even more radical.

\subsection{The Church-Turing-Deutsch principle}

Consider the following
\begin{quote}
{\bf The Church-Turing-Deutsch principle:} Every physical process can be simulated by a universal model computing
device.
\end{quote}
Deutsch \cite{Deutsch} was motivated to state this principle by work of Church \cite{Church} and Turing \cite{Turing}
(actually he gives a stronger and more carefully formulated version).  Deutsch's statement emphasizes the physical
aspect of computation whereas Church and Turing were more interested in mathematical issues (note that, in his
acknowledgements, Deutsch thanks \lq\lq C.\ H.\ Bennett for pointing out to me that the Church-Turing hypothesis has
physical significance"). We can take the widespread successful simulation of any number of physical processes (such as
of cars in a wind tunnel, or of bridges prior to their being built) on a modern classical computer, as evidence of the
truth of this principle. A principle like this would seem to be important since it provides a mechanism for verifying
physical theories. The physical theory tells us how to model physical processes. To verify the physical theory there
needs to be some way of using the theory to simulate the given physical process. However, there is a deeper reason that
this principle is interesting.  This is that it might lead us to say that the universe is, itself, a computer.  Of
course, the CTD principle does not actually imply that. Even though we might be able to simulate a physical process on
a computer, it does not follow that the computation is an accurate reflection of what is happening during that physical
process.  This suggests a stronger principle
\begin{quote}
{\bf  The computational reflection principle:} The behaviour of any physical process is accurately reflected by the
behaviour of an appropriately programmed universal model computing device.
\end{quote}
A proper understanding of this principle requires a definition of what is meant by \lq\lq accurately reflected" (note
that a dictionary definition of the relevant meaning of the word {\it reflect} is to \lq\lq embody or represent in a
faithful or appropriate way" \cite{OEDConcise}).  We will not attempt to provide a precise definition but rather will
illustrated our discussion with examples.  Nevertheless, \lq\lq accurate reflection" would entail that not only is
there the same mapping between inputs and outputs for the physical process and the computation, but also that there is
a mapping between the internal structure of the physical process and the computation.  This relates to ideas of
functional equivalence as discussed by philosophers.

We may think of a universal computer in the Turing model where the program is included in the tape.  But we may also
use the circuit model where the program is represented by a prespecified way of choosing the gates.

It is possible to simulate any quantum system with a finite dimensional Hilbert space (including quantum computers) to
arbitrary accuracy on a classical computer.  In fact, we can even simulate a quantum computer with polynomial space on
a classical computer but, in general, this requires exponential time \cite{simqc}.  We might claim, then, that the CTD
principle holds (though, since this is not exact simulation, we may prefer to withhold judgment).  However, we would be
more reluctant to claim that the CR principle holds since the classical simulation has properties that the quantum
process does not: (i) It is possible to measure the state of the classical computer without effecting its subsequent
evolution; (ii) the exponential time classical computer is much more powerful than a polynomial time quantum computer;
and (iii) the detailed structure of the classical computation will look quite different to that of the quantum process.

\subsection{Physics without state evolution}

The idea of a state which evolves is deeply ingrained in our way of thinking about the world.  But is it a necessary
feature of any physical theory?  This depends what a physical theory must accomplish.  At the very least, {\it a
physical theory must correlate recorded data}.  Data is correlated in the evolving state picture in the following way.
Data corresponding to a given time is correlated by applying the mathematical machinery of the theory to the state at
that given time.  And data corresponding to more than one time is correlated by evolving the state through those given
times, and then applying the mathematical machinery of the theory to the collection of states so generated.  However,
there is no reason to suppose that this is the only way of correlating data taken in different spacetime regions.  In
fact, we have already other pictures.  In GR we solve local field equations.  A solution must simply satisfy the
Einstein field equations and be consistent with the boundary conditions.  We do not need the notion of an evolving
state here - though there are canonical formulations of GR which have a state across space evolving in time. In
classical mechanics we can extremise an action. In this case we consider possible solutions over all time and find the
one that extremizes the action. Again, we do not need to use the notion of an evolving state. In quantum theory we can
use Feynman's sum over histories approach which is equivalent to an evolving state picture but enables us to proceed
without such a picture.  In \cite{Hardy1} the causaloid formalism was developed as a candidate framework for a theory
of QG (though QT can be formulated in this framework). This enables one to calculate directly whether (i) there is a
well defined correlation between data taken from two different spacetime regions and, if there is, (ii) what that
correlation is equal to.  Since this calculation is direct, there is no need to consider a state evolving between the
two regions. The causaloid formalism is, in particular, suited to dealing with the situation where there is no matter
of fact to whether an interval is time-like or not.

\subsection{What is a quantum gravity computer?}\label{whatis}

A quantum gravity computer is a computer for which the particular effects of QG are important.  In this paper we are
interested in the case where we have indefinite causal structure (and, of course, we are assuming that QG will allow
this property).

As we discussed in Sec. \ref{whatcounts}, a computer can be understood to be a physical device having an output that
depends on an input (or program) according to sufficiently well known rules.  The computer occupies a certain region of
spacetime. The input can consist of a number of inputs into the computer distributed across this region, and likewise,
the output can consist of a number of outputs from the computer distributed across the region.  Typically the inputs
are selected (by us) in accordance with some program corresponding to the question we wish to use the computer to find
an answer to. Usually we imagine setting the computer in some initial state (typically, in quantum computing, this
consists of putting all the qubits in the zero state). However, physically this is accomplished by an appropriate
choice of inputs prior to this initial time (for example, we might have a quantum circuit which initializes the state).
Hence, the picture in which we have inputs and outputs distributed across the given region of spacetime is sufficient.
We do not need to also imagine that we separately initialize the computer. This characterization of a computer is
useful for specifying a QG computer since we must be careful using a notion like \lq\lq initial state" when we cannot
rely on having a definite notion of a single time hypersurface in the absence of definite causal structure. The QG
computer itself must be sensitive to QG effects (as opposed to purely quantum or purely general relativistic effects).
To actually build a QG computer we need a theory of quantum gravity because (i) this is the only way to be sure we are
seeing quantum gravity effects and (ii) we need to have known physical laws to use the device as a computer.

In the absence of a theory of QG it is difficult to give an example of a device which will function as a QG computer.
Nevertheless we will give a possible candidate example for the purposes of discussion.  We hope that the essential
features of this example would be present in any actual QG computer.  We imagine, for this example, that our quantum
gravity computer consists of a number of mesoscopic probes of Planck mass (about 20 micrograms) immersed in a
controlled environment of smaller quantum objects (such as photons). There must be the possibility of having inputs and
outputs. The inputs and outputs are distributed across the region of spacetime in which the QG computer operates. We
take this region of spacetime to be fuzzy in the sense that we cannot say whether a particular interval in it is
time-like or space-like. However, we can still expect to be able to set up physical coordinates to label where a
particular input or output is \lq\lq located" in some appropriate abstract space.  For example, imagine that a GPS
system is set up by positioning four satellites around the region. Each satellite emits a signal carrying the time of
its internal clock.  We imagine that the mesoscopic probes can detect these four times thus providing a position
$x\equiv(t_1, t_2, t_3, t_4)$. Each satellite clock will tick and so $x$ is a discrete variable. A given probe will
experience a number of different values of $x$. Assume that each probe can be set to give out a light pulse or not
(denote this by $s=1$ or $s=0$ respectively), and has a detector which may detect a photon or not (denote this by $a=1$
or $a=0$ respectively) during some given short time interval.  Further, allow the value of $s$ to depend on $x$. Thus,
\begin{equation}
s = F(x, n)
\end{equation}
where $n$ labels the probe.  We imagine that we can choose the function $F$ as we like. This constitutes the program.
Thus, the inputs are given by the $s$'s and the outputs by the $a$'s.  We record many instances of the data $(x, n, s,
a)$.  We might like to have more complicated programs where $F$ is allowed to depend on the values of previous outputs
from other probes.  However, we cannot assume that there is fixed causal structure, and so we cannot say, in advance,
what will constitute previous data.  Thus, any program of this nature must \lq\lq physicalize" the previous data by
alowing the probe to emit it as a physical signal, $r$.  If this signal is detected at a probe along with $x$ then it
can form part of the input into $F$.  Thus, we would have
\begin{equation}
s = F(x, n, r)
\end{equation}
At the end of a run of the QG computer, we would have many instances of $(x, n, r, s, a)$.

This is just a possible example of a possible QG computer. We might have the property of indefinite causal structure in
this example since the mesoscopic probes are (possibly) sufficiently small to allow quantum effects and sufficiently
massive to allow gravitational effects.  Penrose's cat \cite{Penrose} consists of exploring the possible gravity
induced breakdown of quantum theory for a Planck mass mirror recoiling (or not) from a photon in a quantum
superposition.

Regardless of whether this is a good example, we will assume that any such computer will collect data of the form $(x,
n, s, a)$ (or $(x, n, r, s, a)$), and that a program can be specified by a function $F(x, n)$ (or $F(x, n, r)$). Whilst
we can imagine more complicated examples, it would seem that they add nothing extra and could, anyway, be accommodated
by the foregoing analysis.  Importantly, although we have the coordinate $x$, we do not assume any causal structure on
$x$.  In particular, there is no need to assume that some function of $x$ will provide a time coordinate - this need
not be a SC.

\section{The causaloid formalism}

\subsection{Analyzing data}

We will now given an abbreviated presentation of the causaloid formalism which is designed for analyzing data collected
in this way and does not require a time coordinate. This formalism was first presented in \cite{Hardy1} (see also
\cite{Hardy2} and \cite{Hardy3} for more accessible accounts). Assume that each piece of data ($(x, n, s, a)$ or $(x,
n, r, s, a)$) once collected is written on a card. At the end of the computation we will have a stack of cards. We will
seek to find a way to calculate probabilistic correlations between the data collected on these cards.  The order in
which the cards end up in the stack does not, in itself, constitute recorded data and consequently will play no role in
this analysis. Since we are interested in probabilities we will imagine running the computation many times so that we
can calculate probabilities as relative frequencies (though, this may not be necessary for all applications of the
computer).  Now we will provide a number of basic definitions in terms of the cards.
\begin{description}
\item[The full pack,] $V$, is the set of all logically possible cards.
\item[The program,] $F$, is the set of all cards from $V$ consistent with a given program $F(x, n, s, a)$ (or $F(x,
n, r, s, a)$).  Note that the set $F$ and the function $F$ convey the same information so we use the same notation,
the meaning being clear from the context.
\item[A stack,] $Y$, is the set of cards collected during a particular run of the computer.
\item[An elementary region,] $R_x$, is the the set of all cards from $V$ having a particular $x$ written on them.
\end{description}
Note that
\begin{equation}
Y \subseteq F \subseteq V
\end{equation}
We will now give a few more definitions in terms of these basic definitions.
\begin{description}
\item[Regions.] We define a composite spacetime region by
\begin{equation}
R_{{\cal O}_1} =\bigcup_{x\in {\cal O}_1 } R_x
\end{equation}
We will often denote this by $R_1$ for shorthand.
\item[The outcome set in region] $R_1$ is given by
\begin{equation}
Y_{R_1} \equiv Y \cap R_1
\end{equation}
This set contains the results seen in the region $R_1$.  It constitutes the raw output data from the computation.  We
will often denote this set by $Y_1$.
\item[The program in region] $R_1$ is given by
\begin{equation}
F_{R_1} \equiv F \cap R_1
\end{equation}
This set contains the program instructions in region $R_1$.  We will often denote it by $F_1$.
\end{description}

\subsection{Objective of the causaloid formalism}\label{objectiveof}

We will consider probabilities of the form
\begin{equation}\label{genprob}
{\rm Prob}(Y_2|Y_1, F_2, F_1)
\end{equation}
This is the probability that we see outcome set $Y_2$ in $R_2$ given that we have procedure $F_2$ in that region and
that we have outcome set $Y_1$ and program $F_1$ in region $R_1$.  Our physical theory must (i) determine whether the
probability is {\it well defined}, and if so (ii) determine its value.  The first step is crucial. Most conditional
probabilities we might consider are not going to be well defined.  For example if $R_1$ and $R_2$ are far apart (in so
much as such a notion makes sense) then there will be other influences (besides those in $R_1$) which determine the
probabilities of outcomes in $R_2$, and if these are not take into account we cannot do a calculation for this
probability.  To illustrate this imagine an adversary.  Whatever probability we write down, he can alter these
extraneous influences so that the probability is wrong.  Conventionally we determine whether a probability is well
defined by simply looking at the causal structure.  However, since we do not have definite causal structure here we
have to be more careful.

To begin we will make an assumption.  Let the region $R$ be big (consisting of most of $V$).
\begin{description}
\item[Assumption 1:] We assume that there is some condition $C$ on $F_{V-R}$ and $Y_{V-R}$ such that all probabilities of
the form
\begin{equation}
{\rm Prob}(Y_R|F_R, C)
\end{equation}
are well defined.
\end{description}
We can regard condition $C$ as corresponding to the setting up and maintenance of the computer.  We will consider only
cases where $C$ is true (when it is not, the computer is broken or malfunctioning).  We will regard region $R$ as the
region in which the computation is performed.  Since we will always be assuming $C$ is true, we will drop it from our
notation. Thus, we assume that the probabilities ${\rm Prob}(Y_R|F_R)$ are well defined.

The probabilities ${\rm Prob}(Y_R|F_R)$ pertain to the global region $R$.  However, we normally like to do physics
by building up a picture of the big from the small.  We will show how this can be done.  We will apply three levels
of {\it physical compression}.  The first applies to single regions (such as $R_1$).  The second applies to
composite regions such as $R_1\cup R_2$ (the second level of physical compression also applies to composite regions
made from three or more component regions).  The first and second levels of physical compression result in certain
matrices. In the third level of physical compression we use the fact that these matrices are related to implement
further compression.

\subsection{First level physical compression}

First we implement {\it first level physical compression}.  We label each possible pair $(Y_{R_1}, F_{R_1})$ in $R_1$
with $\alpha_1$.  We will think of these pairs as describing measurement outcomes in $R_1$ ($Y_{R_1}$ denotes the
outcome of the measurement and $F_{R_1}$ denotes the choice of measurement). Then we write
\begin{equation}
p_{\alpha_1} \equiv {\rm Prob}(Y_{R_1}^{\alpha_1} \cup Y_{R-R_1} |F_{R_1}^{\alpha_1} \cup F_{R-R_1} )
\end{equation}
By Assumption 1, these probabilities are all well defined.  We can think of what happens in region $R-R_1$ as
constituting a generalized preparation of a state in region $R_1$.  We define the state to be that thing represented by
any mathematical object which can be used to calculate $p_{\alpha_1}$ for all $\alpha_1$.  Now, given a generalized
preparation, the $p_{\alpha_1}$'s are likely to be related by the physical theory that governs the system.  In fact we
can just look at linear relationships.  This means that we can find a minimal set $\Omega_1$ such that
\begin{equation}\label{firstlevel}
p_{\alpha_1} = {\bf r}_{\alpha_1}(R_1) \cdot {\bf p}(R_1)
\end{equation}
where the state ${\bf p}(R_1)$ in $R_1$ is given by
\begin{equation}
{\bf p}(R_1)= \left( \begin{array}{c} \vdots \\ p_{l_1} \\ \vdots \end{array} \right) ~~~~ l_1\in \Omega_1
\end{equation}
We will call $\Omega_1$ the fiducial set (of measurement outcomes). Note that the probabilities $p_{l_1}$ need not add
up to 1 since the $l_1$'s may correspond to outcomes of incompatible measurements.  In the case that there are no
linear relationships relating the $p_{\alpha_1}$'s we set $\Omega_1$ equal to the full set of $\alpha_1$'s and then
${\bf r}_{\alpha_1}$ consists of a 1 in position $\alpha_1$ and 0's elsewhere. Hence, we can always write
(\ref{firstlevel}). One justification for using linear compression is that probabilities add in a linear way when we
take mixtures.  It is for this reason that linear compression in quantum theory (for general mixed states) is the most
efficient.  The set $\Omega_1$ will not, in general, be unique. Since the set is minimal, there must exist a set of
$|\Omega_1|$ linearly independent states ${\bf p}$ (otherwise further linear compression would be possible). First
level physical compression for region $R_1$ is fully encoded in the matrix
\begin{equation}
\Lambda_{\alpha_1}^{l_1} \equiv r_{l_1}^{\alpha_1}
\end{equation}
where $r_{l_1}^{\alpha_1}$ is the $l_1$ component of ${\bf r}_{\alpha_1}$. The more physical compression there is
the more rectangular (rather than square) this matrix will be.

\subsection{Second level physical compression}

Next we will implement second level physical compression.  Consider two regions $R_1$ and $R_2$.  Then the state
for region $R_1\cup R_2$ is clearly of the form
\begin{equation}
{\bf p}(R_1\cup R_2)= \left( \begin{array}{c} \vdots \\ p_{k_1k_2} \\ \vdots \end{array} \right) ~~~~ k_1k_2\in
\Omega_{12}
\end{equation}
We can show that it is always possible to choose $\Omega_{12}$ such that
\begin{equation}\label{centralresult}
\Omega_{12} \subseteq \Omega_1\times\Omega_2
\end{equation}
where $\times$ denotes the cartesian product.  This result is central to the causaloid formalism.  To prove
(\ref{centralresult}) note that we can write $p_{\alpha_1\alpha_2}$ as
\begin{eqnarray}
\lefteqn{ {\rm prob}(Y^{\alpha_1}_{R_1}\cup Y^{\alpha_2}_{R_2}\cup Y_{R-R_1-R_2}   |  F^{\alpha_1}_{R_1}\cup
F^{\alpha_2}_{R_2}\cup F_{R-R_1-R_2}) }
\qquad\qquad\qquad\qquad\qquad\qquad \nonumber\\
&=&{\bf r}_{\alpha_1}(R_1)\cdot {\bf p}_{\alpha_2}(R_1)  \nonumber \\
&=&\sum_{l_1\in\Omega_1} r^{\alpha_1}_{l_1}(R_1) p^{\alpha_2}_{l_1}(R_1)   \nonumber \\
&=&\sum_{l_1\in\Omega_1} r^{\alpha_1}_{l_1}(R_1) {\bf r}_{\alpha_2}(R_2)\cdot {\bf p}_{l_1}(R_2) \nonumber \\
&=& \sum_{l_1l_2\in\Omega_1\times\Omega_2} r^{\alpha_1}_{l_1} r^{\alpha_2}_{l_2} p_{l_1l_2} \label{mainproof}
\end{eqnarray}
where ${\bf p}_{\alpha_2}(R_1)$ is the state in $R_1$ given the generalized preparation $(Y^{\alpha_2}_{R_2}\cup
Y_{R-R_1-R_2}, F^{\alpha_2}_{R_2}\cup F_{R-R_1-R_2})$ in region $R-R_1$, and ${\bf p}_{l_1}(R_2)$ is the state in $R_2$
given the generalized preparation $(Y^{l_1}_{R_1}\cup Y_{R-R_1-R_2}, F^{l_1}_{R_1}\cup F_{R-R_1-R_2})$ in region
$R-R_2$, and where
\begin{equation}
p_{l_1l_2} = {\rm prob}(Y^{l_1}_{R_1}\cup Y^{l_2}_{R_2} \cup Y_{R-R_1-R_2} | F^{l_1}_{R_1}\cup F^{l_2}_{R_2} \cup
F_{R-R_1-R_2})
\end{equation}
Now we note from (\ref{mainproof}) that $p_{\alpha_1\alpha_2}$ is given by a linear sum over the probabilities
$p_{l_1l_2}$ where $l_1l_2\in\Omega_1\times\Omega_2$.   It may even be the case that we do not need all of these
probabilities.  Hence, it follows that $\Omega_{12}\subseteq\Omega_1\times \Omega_2$ as required.

Using  (\ref{mainproof}) we have
\begin{eqnarray}
p_{\alpha_1\alpha_2} &=& {\bf r}_{\alpha_1\alpha_2}(R_1\cup R_2)\cdot {\bf p}(R_1\cup R_2) \nonumber \\
                     &=& \sum_{l_1l_2} r^{\alpha_1}_{l_1} r^{\alpha_2}_{l_2} p_{l_1l_2} \nonumber \\
                     &=& \sum_{l_1l_2} r^{\alpha_1}_{l_1} r^{\alpha_2}_{l_2} {\bf r}_{l_1l_2}\cdot {\bf
                              p}(R_1\cup R_2)                                            \nonumber
\end{eqnarray}
We must have
\begin{equation}
{\bf r}_{\alpha_1\alpha_2}(R_1\cup R_2) = \sum_{l_1l_2} r^{\alpha_1}_{l_1} r^{\alpha_2}_{l_2} {\bf
r}_{l_1l_2}(R_1\cup R_2)
\end{equation}
since we can find a spanning set of linearly independent states ${\bf p}(R_1\cup R_2)$. We define
\begin{equation}
\Lambda_{l_1l_2}^{k_1k_2} \equiv r^{l_1l_2}_{k_1k_2}
\end{equation}
where $r^{l_1l_2}_{k_1k_2}$ is the $k_1k_2$ component of ${\bf r}_{l_1l_2}$.  Hence,
\begin{equation}\label{causcomps}
 r^{\alpha_1\alpha_2}_{k_1k_2} = \sum_{l_1l_2} r^{\alpha_1}_{l_1} r^{\alpha_2}_{l_2} \Lambda_{l_1l_2}^{k_1k_2}
\end{equation}
This equation tells us that if we know $\Lambda_{l_1l_2}^{k_1k_2}$ then we can calculate ${\bf
r}_{\alpha_1\alpha_2}(R_1\cup R_2)$ for the composite region $R_1\cup R_2$ from the corresponding vectors ${\bf
r}_{\alpha_1}(R_1)$ and ${\bf r}_{\alpha_2}(R_2)$ for the component regions $R_1$ and $R_2$.  Hence the matrix
$\Lambda_{l_1l_2}^{k_1k_2}$ encodes the second level physical compression (the physical compression over and above
the first level physical compression of the component regions).  We can use $\Lambda_{l_1l_2}^{k_1k_2}$ to define a
new type of product - the causaloid product - denoted by $\otimes^\Lambda$.
\begin{equation}
{\bf r}_{\alpha_1\alpha_2}(R_1\cup R_2) = {\bf r}_{\alpha_1}(R_1)\otimes^{\Lambda} {\bf r}_{\alpha_2}(R_2)
\end{equation}
where the components are are given by (\ref{causcomps}).

We can apply second level physical compression to more than two regions.  For three regions we have the matrices
\begin{equation}
\Lambda_{l_1l_2l_3}^{k_1k_2k_3}
\end{equation}
and so on.

\subsection{Third level physical compression}

Finally, we come to third level physical compression.  Consider all the compression matrices we pick up for
elementary regions $R_x$ during first and second level compression.  We have
\begin{equation}
\left(
\begin{array}{ll}
\Lambda_{\alpha_x}^{l_x}                        & \text{for all}~ x \in {\cal O}_R  \\  \\
\Lambda_{l_xl_{x'}}^{k_xk_{x'}}                 & \text{for all}~ x, x' \in{\cal O}_R \\  \\
\Lambda_{l_xl_{x'}l_{x''}}^{k_xk_{x'}k_{x''}}   & \text{for all}~ x, x',x'' \in{\cal O}_R \\  \\
 ~~~\vdots                                      &~~~ \vdots
\end{array} \right)
\end{equation}
where ${\cal O}_R$ is the set of $x$ in region $R$.  Now, these matrices themselves are likely to be related by the
physical theory.  Consequently, rather than specifying all of them separately, we should be able to specify a subset
along with some rules for calculating the others
\begin{equation}
{\bf \Lambda}\equiv ( \text{subset of }~\Lambda{\rm 's} ; \text{RULES} )
\end{equation}
We call this mathematical object the {\it causaloid}.   This third level of physical compression is accomplished by
identities relating the higher order $\Lambda$ matrices (those with more indices) to the lower order ones.  Here are
some examples from two families of such identities.  The first family uses the property that when $\Omega$ sets
multiply so do $\Lambda$ matrices.
\begin{equation}\label{identity1}
\Lambda_{l_x\cdots l_{x'}l_{x''}\cdots l_{x'''}}^{k_x\cdots k_{x'}k_{x''}\cdots k_{x'''}} = \Lambda_{l_x\cdots
l_{x'}}^{k_x\cdots k_{x'}}\Lambda_{l_{x''}\cdots l_{x'''}}^{k_{x''}\cdots k_{x'''}}~~~\text{if}~~~ \Omega_{x\cdots
x' x'' \cdots x'''} = \Omega_{x\cdots x'} \times \Omega_{x''\cdots x'''}
\end{equation}
The second family consists of identities from which $\Lambda$ matrices for composite regions can be calculated from
some pairwise matrices (given certain conditions on the $\Omega$ sets).  The first identity in this family is
\begin{equation}\label{identity2}
\Lambda_{l_1l_2l_3}^{k_1k_2k_3}=\sum_{k'_2\in\Omega_{2\not{\,3}}}\Lambda_{l_1k'_2}^{k_1 k_2}
\Lambda_{l_2l_3}^{k'_2k_3} ~~~{\rm if}~~~ \Omega_{123}= \Omega_{12}\times \Omega_{\not{\,2} 3} ~~~{\rm and}~~~
\Omega_{23}=\Omega_{2\not{\,3}}\times\Omega_{\not{\,2} 3}
\end{equation}
where the notation $\Omega_{\not{\,2} 3}$ means that we form the set of all $k_3$ for which there exists
$k_2k_3\in\Omega_{23}$.  The second identity in this family is
\begin{equation}\label{identity2b}
\Lambda_{l_1l_2l_3l_4}^{k_1k_2k_3k_4}=\sum_{k'_2\in\Omega_{2\not{\,3}}, k'_3\in\Omega_{3\not{\,4}}}
\Lambda_{l_1k'_2}^{k_1 k_2} \Lambda_{l_2k'_3}^{k'_2k_3} \Lambda_{l_3l_4}^{k'_3k_4} ~~~{\rm if}~~
\begin{array}{l}
\Omega_{1234}= \Omega_{12}\times \Omega_{\not{\,2} 3} \times \Omega_{\not{\,3} 4} \\
\Omega_{23}=\Omega_{2\not{\,3}}\times\Omega_{\not{\,2} 3}  \\
 \Omega_{34}=\Omega_{3\not{\,4}}\times\Omega_{\not{\,3} 4}
\end{array}
\end{equation}
and so on.   These identities are sufficient to implement third level physical compression for classical and quantum
computers. However, we will probably need other identities to implement third level physical compression for a QG
computer. The task of fully characterizing all such identities, and therefore of fully characterizing third level
physical compression, remains to be completed.

\subsection{Classical and quantum computers in the causaloid formalism}\label{classicalquantum}

\begin{figure*}[t]
\centering
{\includegraphics{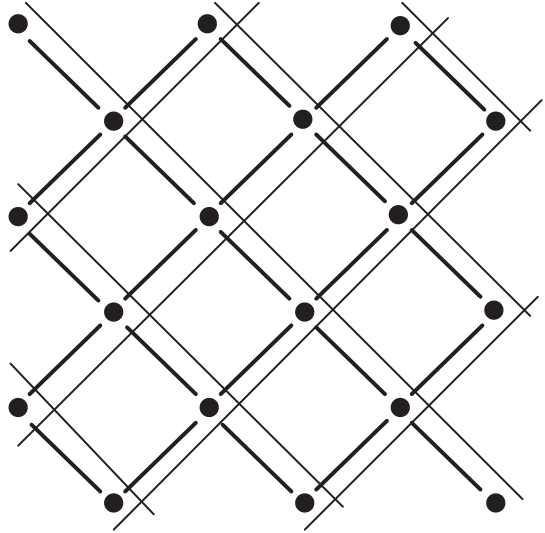}} \caption{\small This figure shows a number of pairwise interacting
(qu)bits. The (qu)bits travel along the paths indicated by the thin lines and interact at the nodes. At each node we
can choose a gate.}\label{Fig1}
\end{figure*}

Since third level compression has been worked out for classical and quantum computers we should say a little about this
here (see \cite{Hardy1} and \cite{Hardy2} for more details).  Consider a classical (quantum) computer which consists of
pairwise interacting (qu)bits. This is sufficient to implement universal classical (quantum) computation.  This
situation is shown in Fig.\ 1.  Each (qu)bit is labeled by $i, j, \dots$ and is shown by a thin line. The nodes where
the (qu)bits meet are labeled by $x$. Adjacent nodes (between which a (qu)bit passes) have a link.  We call this
diagram a {\it causaloid diagram}. At each node we have a choice, $s$, of what gate to implement. And then there may be
some output, $a$, registered at the gate itself (in quantum terms this is both a transformation and a measurement). We
record $(x, s, a)$ on a card. The program is specified by some function $s=F(x)$.  We can use our previous notation.
Associated with each $(x, s, a)$ at each gate is some ${\bf r}_{\alpha_x}$. It turns out that there exists a choice of
fiducial measurement outcomes at each node $x$ which break down into separate measurement outcomes for each of the two
(qu)bits passing through that node. For these measurements we can write $l_x \equiv l_{xi}l_{xj}$ where $l_{xi}$ labels
the fiducial measurements on (qu)bit $i$ and $l_{xj}$ labels the fiducial measurements on the other (qu)bit $j$.  All
$\Omega$ sets involving different (qu)bits factorize  as do all $\Omega$ sets involving non-sequential clumps of nodes
on the same (qu)bit and so identity (\ref{identity1}) applies in these cases.  For a set of sequential nodes the
$\Omega$ sets satisfy the conditions for (\ref{identity2}, \ref{identity2b}) and related identities to hold. This means
that it is possible to specify the causaloid for a classical (quantum) computer of pairwise interacting (qu)bits by
\begin{equation}
{\bf \Lambda}=\left(\{\Lambda_{\alpha_x}^{l_{xi}l_{xj}} ~\forall~ x\},
\{\Lambda_{l_{xi}l_{x'i}}^{k_{xi}k_{x'i}}~\forall~ \text{adjacent} ~x,x' \} ; \left\{  \begin{array}{l} \text{clumping
method}\\ \text{causaloid diagram}\end{array}\right\} \right)
\end{equation}
where the \lq\lq clumping method" is the appropriate use of the identities (\ref{identity1}, \ref{identity2},
\ref{identity2b}) and related identities to calculate general $\Lambda$ matrices.   The causaloid diagram is also
necessary so we know how the nodes are linked up and how the (qu)bits move.  There is quite substantial third level
compression.  The total number of possible $\Lambda$ matrices is exponential in the number of nodes but the number
of matrices required to specify the causaloid is only linear in this number.   There may be simple symmetries which
relate the matrices living on each node and each link.  In this case there will be even further compression.

\subsection{Using the causaloid formalism to make predictions}

We can use the causaloid to calculate any ${\bf r}$ vector for any region in $R$.   Using these we can calculate
whether any  probability of the form (\ref{genprob}) is well defined, and if so, what it is equal to.   To see this
note that, using Bayes rule,
\begin{equation}
p\equiv {\rm Prob}(Y_1^{\alpha_1}|Y_2^{\alpha_2}, F_1^{\alpha_1}, F_2^{\alpha_2})= \frac{{\bf
r}_{\alpha_1\alpha_2}(R_1\cup R_2) \cdot {\bf p}(R_1\cup R_2)}{ \sum_{\beta_1}{\bf r}_{\beta_1\alpha_2}(R_1\cup R_2)
\cdot {\bf p}(R_1\cup R_2)}
\end{equation}
where $\beta_1$ runs over all $(Y_1, F_1)$ consistent with $F_1=F_1^{\alpha_1}$ (i.e. all outcomes consistent with the
program in region $R_1$).  For this probability to be well defined it must be independent of what happens outside
$R_1\cup R_2$.  That is, it must be independent of the state ${\bf p}(R_1\cup R_2)$.  Since there exists a spanning set
of linearly independent such states, this is true if and only if
\begin{equation}\label{parallelness}
{\bf r}_{\alpha_1\alpha_2}(R_1\cup R_2) ~~\text{is parallel to} ~~ \sum_{\beta_1}{\bf r}_{\beta_1\alpha_2}(R_1\cup R_2)
\end{equation}
This, then, is the condition for the probability to be well defined. In the case that this condition is satisfied
then the probability is given by the ratio of the lengths of these two vectors. That is by
\begin{equation}\label{pgivenby}
{\bf r}_{\alpha_1\alpha_2}(R_1\cup R_2) = p\sum_{\beta_1}{\bf r}_{\beta_1\alpha_2}(R_1\cup R_2)
\end{equation}
It might quite often turn out that these two vectors are not exactly parallel.  So long as they are still quite
parallel we can place limits on $p$.  Set
\begin{equation}
{\bf v}\equiv {\bf r}_{\alpha_1\alpha_2}(R_1\cup R_2) ~~~\text{and}~~~ {\bf u}\equiv \sum_{\beta_1}{\bf
r}_{\beta_1\alpha_2}(R_1\cup R_2)
\end{equation}
Define ${\bf v}^\parallel$ and ${\bf v}^\perp$ as the components of ${\bf v}$ parallel and perpendicular to ${\bf
u}$ respectively. Then it is easy to show that
\begin{equation}\label{pbounded}
\frac{|{\bf v}^\parallel|}{|{\bf u}|} - \frac{|{\bf v}^\perp|}{|{\bf v}|\cos\phi} \leq p \leq \frac{|{\bf
v}^\parallel|}{|{\bf u}|} + \frac{|{\bf v}^\perp|}{|{\bf v}|\cos\phi}
\end{equation}
where $\phi$ is the angle between ${\bf v}$ and ${\bf v}^\perp$ (we get these bounds using $|{\bf v}\cdot{\bf
p}|\leq |{\bf u}\cdot{\bf p}|$).

\subsection{The notion of state evolution in the causaloid formalism}\label{thenotion}

In setting up the causaloid formalism we have not had to assume that we can have a state which evolves with respect
to time.   As we will see, it is possible to reconstruct an evolving state even though this is looks rather
unnatural from point of view of the causaloid formalism.  However, this reconstruction depends on Assumption 1 of
Sec.\ \ref{objectiveof} being true.  It is consistent to apply the causaloid formalism even if Assumption 1 does
not hold. In this case we cannot reconstruct an evolving state.

We choose a nested set of spacetime regions $R_t$ where $t=0$ to $T$ for which
\begin{equation}
R_0 \supset R_1 \supset R_2 \dots \supset R_T
\end{equation}
where $R_0=R$ and $R_T$ is the null set.  We can think of $t$ as a \lq\lq time" parameter and the region $R_t$ as
corresponding to all of $R$ that happens \lq\lq after" time $t$.   For each region $R_t$ we can calculate the state,
${\bf p}(t)\equiv {\bf p}(R_t)$, given some generalized preparation up to time $t$ (that is in the region $R-R_t$). We
regard ${\bf p}(t)$ as the state at time $t$. It can be used to calculate any probability after time $t$ (corresponding
to the region $R_t$) and can therefore be used to calculate probabilities corresponding to the region $R_{t+1}$ since
this is nested inside $R_t$. Using this fact it is easy to show that the state is subject to linear evolution so that
\begin{equation}
{\bf p}(t+1) = Z_{t, t+1} {\bf p}(t)
\end{equation}
where $Z_{t, t+1}$ depends on $Y_{R_t-R_{t+1}}$ and $F_{R_t-R_{t+1}}$.

Thus, it would appear that, although we did not use the idea of an evolving state in setting up the causaloid
formalism, we can reconstruct a state that, in some sense, evolves.  We can do this for {\it any} such nested set
of regions.  There is no need for the partitioning to be generated by a foliation into spacelike hypersurfaces and,
indeed, such a foliation will not exist if the causal structure is indefinite. This evolving state is rather
artificial - it need not correspond to any physically motivated \lq\lq time".

There is a further reason to be suspicious of an evolving state in the causaloid formalism.  To set up this formalism
it was necessary to make Assumption 1 (in Sec. \ref{objectiveof}).   It is likely that this assumption will not be
strictly valid in a theory of QG.  However, we can regard this assumption as providing scaffolding to get us to a
mathematical framework. It is perfectly consistent to suppose that this mathematical framework continues to be
applicable even if Assumption 1 is dropped.  Thus it is possible that we can define a causaloid and then use the
causaloid product and (\ref{parallelness}, \ref{pgivenby}) to calculate whether probabilities are well defined and, if
so, what these probabilities are equal to.  In so doing we need make no reference to the concept of state. In
particular, since we cannot suppose that all the probabilities ${\rm prob}(Y_R|F_R, C)$ are well defined, we will not
be able to force an evolving state picture.   The causaloid formalism provides us with a way of correlating inputs and
outputs across a region of space time even in the absence of the possibility of an evolving state picture.

\section{Computation in the light of the causaloid formalism}

\subsection{Gates}

In the standard circuit model a computer is constructed out of gates selected from a small set of possible gates. The
gates are distributed throughout a spacetime region in the form of a circuit.  Hence we have a number of spacetime
locations (label them by $x$) at which we may place a gate.  At each such location we have a choice of which gate to
select. The gates are connected by \lq\lq wires" along which (qu)bits travel.  This wiring represents the causal
structure of the circuit. Since the wiring is well defined, causal structure cannot be said to be indefinite.  In fact
in classical and quantum computers we can work with a fixed wiring and vary only the choice of gates.  The wires can
form a diamond grid like that shown in Fig.\ 1. Where the wires cross two (qu)bits can pass through a gate.  As long as
we have a sufficient number of appropriate gates we can perform universal computation.  In Sec. \ref{classicalquantum}
we outlined how to put this situation into the causaloid formalism.

In the causaloid model we have spacetime locations labeled by $x$.  At each $x$ we have a choice of setting $s$. This
choice of setting can be regarded as constituting the choice of a gate.  Since we may have indefinite causal structure
we will not be able to think in terms of \lq\lq wiring" as such.  However information about the causal connections
between what happens at different $x$'s is given by the $\Lambda$ matrices which can be calculated from the causaloid.
For example the matrix $\Lambda_{l_xl_{x'}}^{k_xk_{x'}}$ tells us about the causal connection between $x$ and $x''$ by
quantifying second level compression.  Thus, the matrices associated with second level compression (which can be
deduced from the causaloid, ${\bf \Lambda}$) play the role of wiring.  Since we do not have wires we cannot necessarily
think in terms of (qu)bits moving between gates.  Rather, we must think of the gates as being immersed in an amorphous
interconnected sea quantitatively described by the causaloid.  In the special case of a classical or quantum computer
we will have wiring and this can be deduced from ${\bf\Lambda}$.

Typically, in computers, we restrict the set of gates we employ. Thus, assume that we restrict to $s\in\{ s_1, s_2,
\dots, s_N\}\equiv S$ where $S$ is a subset of the set, $S_I$, of all possible $s$.  Then a computer is defined by the
pair
\begin{equation}
\{ {\bf\Lambda}, S\}   \text{     where    } S\subset S_I
\end{equation}
The program for this computer is given by some function like  $s=F(x, n)$ (or $s=F(x, n, r)$) from Sec.\ \ref{whatis}
where $s\in S$.  This is a very general model for computation.  Both classical and quantum computers can be described
in this way as well as computers with indefinite causal structure.

\subsection{Universal computation}

Imagine we have a class of computers.  A universal computer for this class is a member of the class which can be used
to simulate any computer in the class if it is supplied with an appropriate program.  For example, a universal Turing
machine can be used to simulate an arbitrary Turing machine.  This is done by writing the program for the Turing
machine to be simulated into the first part of the tape that is fed into the universal Turing machine. It follows from
their definition that universal computers can simulate each other.

Given a causaloid, ${\bf\Lambda}$ and some integer $M$ we can generate an interesting class of computers - namely the
class $C_{\bf\Lambda}^M$ defined as the class of computers $\{ {\bf\Lambda}, S\}$ for all $S\subset S_I$ such that
$|S|\leq M$.  We will typically be interested in the case that $M$ is a fairly small number (less than 10 say). The
reason for wanting $M$ to be small is that usually we imagine computations being constructed out of a small set of
basic operations.

We can then ask whether there exist any universal computers in this class.  We will say that the computer $\{
{\bf\Lambda}, S_U\}$ with $|S_U|\leq M$ is universal for the class $C_{\bf\Lambda}^M$ if we can use it to simulate an
arbitrary computer in this class. This means that there must exist a simple map from inputs and outputs of the
universal computer to inputs and outputs (respectively) of the computer being simulated such that the probabilities are
equal (or equal to within some specified accuracy).  We will then refer to $S_U$ as a universal set of gates.

If we choose the causaloid ${\bf\Lambda}$ of classical or quantum theory discussed in Sec.\ \ref{classicalquantum} then
it is well established that there exist universal computers for small $M$.  This is especially striking in the quantum
case since there exist a infinite number of gates which cannot be simulated by probabilistic mixtures of other gates.
One way to understand how this is possible in the classical and quantum cases is the following.  Imagine that we want
to simulate $\{ {\bf\Lambda}, S\}$ with $\{ {\bf\Lambda}, S_U\}$.  We can show that any gate in the set $S$ can be
simulated to arbitrary accuracy with some number of gates from the set $S_U$.  Then we can coarse-grain on the diamond
grid to larger diamonds which can have sufficient gates from $S_U$ to simulate an arbitrary gate in $S$.  In
coarse-graining in this way we do not change in any significant way the nature of the causal structure.  Thus we can
still link these coarse-grained diamonds to each other in such a way that we can simulate $\{ {\bf\Lambda}, S\}$.  This
works because, in classical and quantum theory, we have definite causal structure which has a certain scale invariance
property as we coarse-grain.

However, if we start with a class of computers $C_{\bf\Lambda}^M$ generated by a causaloid, ${\bf\Lambda}$, for which
there is indefinite causal structure, then we do not expect this scale invariance property under coarse-graining.  In
particular, we would expect, as we go to larger diamonds, that the causal structure will become more definite.  Hence
we may not be able to arrange the same kind of causal connection between the simulated versions of the gates in $S$ as
between the original versions of these gates.  Hence, we cannot expect that the procedure just described for simulation
in the classical and quantum case will work in the case of a general causaloid.

This suggests that the concept of universal computation is may not be applicable in QG.  However the situation is a
little more subtle.  The classical physics that is required to set up classical computation should be a limiting case
of any theory of QG. If a given causaloid, ${\bf \Lambda}$, corresponds to QG then we expect that it is possible to use
this to simulate a universal classical computer if we coarse-grain to a classical scale.  We can also build random
number generator since we have probabilistic processes (since QT is also a limiting case).  This suggests a way to
simulate (in some sense of the word) a general QG computer in the class corresponding to ${\bf \Lambda}$.  We can use
the classical computer to calculate whether probabilities are well defined and, if so, what they are equal to arbitrary
accuracy from the causaloid by programming in the equations of the causaloid formalism.  We can then use the random
number generator to provide outputs with the given probabilities thus simulating what we would see with a genuine QG
computer. We might question whether this is genuine simulation since there will not necessarily be a simple
correspondence between the spacetime locations of these outputs in the simulation and the outputs in the actual QG
computation. In addition, in simulating the classical computer from the quantum gravitational ${\bf\Lambda}$, we may
need a gate set $S$ with very large $M$.  Nevertheless, one might claim that the Church Turing Deutsch principle is
still true. However, it seems that the computational reflection principle is under considerable strain. In particular,
the classical simulation would have definite causal structure unlike the QG computer.  But also the detailed causal
structure of the classical simulation would look quite different from that of the QG computer it simulates.  There may
also be computational complexity issues.  With such issues in mind we might prefer to use the QG causaloid to simulate
a universal quantum computer (instead of a universal classical computer) and then use this to model the equations of
the causaloid formalism to simulate the original causaloid.  This may be quicker than a classical computer.  However,
the computational power of a QG computer may go significantly beyond that of a quantum computer (see Sec.\
\ref{willQG}).

If the computational reflection principle is undermined for QG processes then we may not be able to think that the
world is, itself, a computational process.  Even if we widen our understanding of what we mean by computation, it is
possible that we will not be able to define a useful notion of a universal computer that is capable of simulating all
fundamental quantum gravitational processes in a way that accurately reflects what is happening in the world.  This
would have an impact on any research program to model fundamental physics as computation (such as that of Lloyd
\cite{Lloyd}) as well as having wider philosophical implications.

\subsection{Will quantum gravity computers have greater computational power than quantum computers?}\label{willQG}

Whether or not we can define a useful notion of universal QG computation, it is still possible that a QG computer will
have greater computational power than a quantum computer (and, therefore, a classical computer).  Are there any reasons
for believing this?

Typically we are interested in how computational resources scale with the input size for a class of problems.  For
example we might want to factorize a number.  Then the input size is equal to the number of bits required to represent
this number.  To talk about computational power we need to a way of measuring resources.  Computer scientists typically
make much use of SPACE and TIME as separate resources. TIME is equal to the number of steps required to complete the
calculation and SPACE is equal to the maximum number of (qu)bits required.  Many complexity classes have been defined.
However, of most interest is the class $P$ of decision problems for which TIME is a polynomial function of the size of
the input on a classical computer (specifically, a Turing machine). Most simple things like addition, multiplication,
and division, are in $P$. However factorization is believed not to be. Problems in P are regarded as being easy and
those which are not in $P$ are regarded as being hard. Motivated by the classical case, $BQP$ is the class of decision
problems which can be solved with bounded error on a quantum computer in polynomial time. Bounded error means that the
error must be, at most, $1/3$.  We need to allow errors since we are dealing with probabilistic machines. However, by
repeating the computation many times we can increase our certainty whilst still only requiring only polynomial time.

In QG computation with indefinite causal structure we cannot talk about SPACE and TIME as separate resources.  We can
only talk of the SPACETIME resources required to complete a calculation.  The best measure of the spacetime resources
is the number of locations $x$ (where gates are chosen) that are used in the computation. Thus, if we have $x\in{\cal
O}$ for a computation then ${\rm SPACETIME}=|{\cal O}|$.

In standard computation, the SPACE used by a computer with polynomial TIME is, itself, only going to be at most
polynomial in the input size (since, in the computational models used by computer scientists, SPACE can only increase
as a polynomial function of the number of steps).  Hence, if a problem is in P then SPACETIME will be a polynomial
function of the input size also.  Hence, we can usefully work with SPACETIME rather than TIME as our basic resource.

We define the class of problems $BP_{\{ {\bf \Lambda}, S \} }$ which can be solved with bounded error on the computer
 $\{ {\bf \Lambda},S\}$ in polynomial SPACETIME. The interesting question,
then, is whether there are problems which are in $BP_{\{ {\bf \Lambda}, S \} }$ but not in BQP for some appropriate
choice of computer $\{ {\bf \Lambda}, S \} $.  The important property that a QG computer will have that is not
possessed by a  quantum (or classical) computer is that we do not have fixed causal structure. This means that, with
respect to any attempted foliation into spacelike hypersurfaces, there will be backward in time influences.  This
suggests that a QG computer will have some  {\it insight} into its future state (of course, the terminology is awkward
here since we do not really have a meaningful notion of \lq\lq future").  It is possible that this will help from a
computational point of view.

A different way of thinking about this question is to ask whether a QG computer will be hard to simulate on a quantum
computer. Assuming, for the sake of argument, that Assumption 1 is true then, as seen in Sec.\ \ref{thenotion}, we can
force an evolving state point of view (however unnatural this may be).  In this case we can simulate the QG computer by
simulating the evolution of ${\bf p}(t)$ with respect to $t$. However, this is likely to be much harder when there is
not the kind of causal structure with respect to $t$ which we would normally have if $t$ was a physically meaningful
time coordinate. In the classical and quantum cases we can determine the state at time $t$ by making measurements at
time $t$ (or at least in a very short interval about this time). Hence, to specify the state, ${\bf p}(t)$, we need
only list probabilities pertaining to the time-slice $R_t-R_{t+1}$ rather than all of $R_t$.  The number of
probabilities required to specify ${\bf p}(t)$ (i.e. the number of entries in this vector) is therefore much smaller
than it might be if we needed to specify probabilities pertaining to more of the region $R_t$. If, however, we have
indefinite causal structure, then we cannot expect to have this property.  Hence the state at time $t$ may require many
more probabilities for its specification. This is not surprising since the coordinate $t$ has no natural meaning in
this case.   Hence, it is likely that we will require much greater computational power to simulate the evolution of
${\bf p}(t)$ simply because we will have to store more probabilities at each stage of the evolution. Hence we can
expect that it will be difficult to simulate a QG computer on a quantum computer. However, an explicit model is
required before we can make a strong claim on this point.

\section{Conclusions}

It is likely that a theory of quantum gravity will have indefinite causal structure.  If this is the case it will have
an impact on the theory of computation since, when all is said and done, computers are physical machines.  We might
want to use such QG effects to implement computation.  However, if there is no definite causal structure we must depart
from the usual notion of a computation as corresponding to taking a physical machine through a time ordered sequence of
steps - a QG computer will likely not be a step computer.  We have shown how, using the causaloid formalism, we can set
up a mathematical framework for computers that may not be step computers.  In this framework we can represent a
computer by the pair $\{ {\bf \Lambda}, S \} $.  Classical and quantum computers can be represented in this way.

We saw that the notion of universal computation may be undermined since the nature of the causal structure is unlikely
to be invariant under scaling (the fuzzyness of the indefinite causal structure is likely to go away at large enough
scales). If this is true then it will be difficult to make the case that the universe is actually a computational
process.

It is possible that the indefinite causal structure will manifest itself as a computational resource allowing quantum
gravity computers to beat quantum computers for some tasks.

An interesting subject is whether general relativity computers will have greater computational powers. There has been
some limited investigation of the consequences of GR for computation for static spacetimes (see \cite{Hogarth1,
Hogarth2} and \cite{Pitowsky}).  General relativity has not been put into the causaloid framework.  To explore the
computational power of GR we would need to put it into an operational framework of this nature.

The theory of quantum gravity computation is interesting in its own right.  Thinking about quantum gravity from a
computational point of view may shed new light on quantum gravity itself - not least because thinking in this way
forces operational clarity about what we mean by inputs and outputs.  Thinking about computation in the light of
indefinite causal structure may shed significant light on computer science - in particular it may force us to loosen
our conception of what constitutes a computer even further than that already forced on us by quantum computation. Given
the extreme difficulty of carrying out quantum gravitational experiments, however, it is unlikely that we will see
quantum gravity computers any time soon.

We have investigated the issue of QG computers in the context of the causaloid framework. This is a candidate framework
for possible theories of QG within which we can use the language of inputs and outputs and can model indefinite causal
structure (a likely property of QG).  The main approaches to QG include String Theory \cite{ST}, Loop Quantum Gravity
\cite{LQG1,LQG2, LQG3}, Causal Sets \cite{CS}, and Dynamical Triangulations \cite{DT}.  These are not formulated in a
way that it is clear what would constitute inputs and outputs as understood by computer scientists.  Aaronson provides
an interesting discussion of some of these approaches and the issue of quantum gravity computation \cite{Aaronson}.  He
concludes that it is exactly this lack of conceptual clarity about what would constitute inputs and outputs that
prohibits the development of a theory of quantum gravity computation.   Whilst the causaloid formalism does not suffer
from this problem, it does not yet constitute an actual physical theory.  It is abstract and lacks physical constants,
dimensionalful quantities, and all the usual hallmarks of physics that enable actual prediction of the values of
measurable quantities.

Issues of computation in the context of quantum gravity have been raised by Penrose \cite{Penrosenewmind,
Penroseshadows}. He has suggested that quantum gravitational processes may be non-computable and that this may help to
explain human intelligence.  In this paper we have chosen to regard quantum gravitational processes as allowing us to
define a new class of computers which may have greater computational powers because they may be able to harness the
indefinite causal structure as a computational resource.  It is likely that QG computers, as understood in this paper,
can be simulated by both classical and quantum computers so they will not be able to do anything that is non-computable
from the point of view of classical and quantum computation.   However, it may require incredible classical or quantum
resources to simulate a basic QG computational process.  Further the internal structure of a QG computation will most
likely be very different to that of any classical or quantum simulation.  Hence, the \lq\lq thought process" on a QG
computer may be very different to that of a classical or quantum computer in solving the same problem and so, in spirit
if not in detail, the conclusions of this paper may add support to Penrose's position.  Of course, QG computation can
only be relevant to the human brain if it can be shown that the particular effects of QG can be resident there
\cite{Tegmark, Penroseshadows}.

\vspace{6mm}

\noindent{\Large\bf Dedication}

\vspace{6mm}

\noindent It is a great honour to dedicate this paper to Abner Shimony whose ideas permeate the field of the
foundations of quantum theory.  Abner has taught us the importance of metaphysics in physics.   I hope that not only
can metaphysics drive experiments (Abner's \lq\lq experimental metaphysics") but that it can also drive theory
construction.

\end{document}